\documentclass[a4paper,12pt]{article}
\usepackage{epsfig}
\usepackage[dvips,usenames]{color}
\usepackage{graphicx}
\usepackage{amssymb,amsmath}

\newlength{\dinwidth}
\newlength{\dinmargin}
\begin{document}
\title{Higgs boson decays and production in the left-right twin Higgs model}
\bigskip
\author{Yao-Bei Liu$^{1}$, Xue-Lei Wang$^{2}$\\
{\small 1: Henan Institute of Science and Technology, Xinxiang
453003, P.R.China}
\thanks{E-mail:liuyaobei@sina.com}\\
{\small 2: College of Physics and Information Engineering,} \\
{\small Henan Normal University, Xinxiang 453007, P.R.China}\\
  }\maketitle
%\date{today}
\begin{abstract}
The left-right twin Higgs model predicts one neutral Higgs boson
$\phi_{0}$ and it acquires mass $m_{\phi_{0}}\sim \mu_{r}$ with the
$\mu$ term, which can be lighter than half the SM-like Higgs boson
mass in a portion of parameter space. Thus, the SM-like Higgs boson
$h$ can dominantly decay into a pair of light neutral Higgs bosons
especially when $m_{h}$ is below the $WW$ threshold. First, we
examine the branching ratios of the SM-like Higgs boson decays and
find that the new decay mode $h\rightarrow \phi_{0}\phi_{0}$ is
dominant for the case of $m_{h}>2m_{\phi_{0}}$. Then we study the
production via gluon fusion followed by the decay into two photons
or two weak gauge bosons and found that the production rate can be
significantly suppressed for some part of parameter space. Finally,
we comparatively study the process $\gamma\gamma\rightarrow h
\rightarrow b\bar{b}$ at ILC in the cases of $m_{h}>2m_{\phi_{0}}$
and $m_{h}<2m_{\phi_{0}}$, respectively. We find that these
predictions can significantly deviated from the SM predictions,
e.g., the gluon-gluon fusion channel, in the cases of
$m_{h}>2m_{\phi_{0}}$ and $m_{h}<2m_{\phi_{0}}$, can be suppressed
by about $80\%$ and $45\%$, respectively. Therefor, it is possible
to probe the left-right twin Higgs model via these Higgs boson
production processes at the LHC experiment or in the future ILC
experiment.
\end{abstract}
\newpage
\noindent{\bf I. Introduction}~~\\
\indent The Higgs boson is the last ingredient of the standard model
(SM) to be probed at experiments. The precision electroweak
measurement
  data and direct searches suggest that the Higgs boson must be relatively light and its mass should
 be roughly in the range of 114.4 GeV$\sim$186 GeV at $95\%$ C.L.
 \cite{lep}, but the SM suffers of the so-called hierarchy problem \cite{hierarchy}, which is due to the presence of quadratic divergences
 in the loop processes for the scalar Higgs boson self-energy. Therefore, the standard model
 with a light Higgs boson can be viewed as the low-energy effective approximation of a fundamental
 theory. A wide variety of models have been introduced to address
 electroweak symmetry breaking (EWSB) and the hierarchy problem:
 supersymmetry\cite{sup}, large extra-dimensions\cite{led}, topcolor models\cite{topclor}, and little Higgs
 models\cite{little} et al.\\
 \indent Recently, the twin Higgs mechanism is proposed as a solution to
the little hierarchy problem \cite{ly-1,ly-2,ly-3}. Instead of
protecting the Higgs mass from receiving large radiative corrections
by using several approximate global symmetries, twin Higgs theories
use a discrete symmetry in combination with an approximate global
symmetry to eliminate the quadratic divergences that arise at loop
level. Together with the gauge symmetries of the model, the discrete
symmetry mimics the effect of a global symmetry, thus stabilizing
the Higgs mass. The twin Higgs mechanism can be implemented in
left-right models with the discrete symmetry being identified with
left-right symmetry \cite{ly-2}. In the left-right twin Higgs(LRTH)
model, the leading quadratically divergent contributions of the SM
gauge bosons to the Higgs boson mass are canceled by the loop
involving the new gauge bosons, while those for the top quark can be
canceled by the contributions from a heavy top quark. Besides, the
other Higgs particles acquire large masses not only at quantum level
but also at tree level. The phenomenology of the LRTH model are
widely discussed in literature \cite{Hock,hock1,su,yl}, and
constraints on LRTH model parameters are studied in \cite{loop}. The
LRTH model is also expected to give new significant signatures in
future high energy colliders and studied in references
\cite{liu}.\\
\indent Besides the SM-like Higgs boson $h$, there are two
additional neutral Higgs bosons in the LRTH model, which are
$\hat{h}_{2}^{0}$ and $\phi_{0}$. The neutral Higgs boson
$\hat{h}_{2}^{0}$ could be a good dark matter candidate\cite{su}.
The light neutral Higgs boson $\phi_{0}$ is a pseudoscalar and
charged under the spontaneously broken $SU(2)_{R}$. Its mass is
determined by $\mu_{r}$ that can be anything below the scale $f$.
Here we consider another possibility, in which the mass
$m_{\phi_{0}}< m_{h}/2$. Therefore, in addition to the SM decay
channels, the Higgs boson can then decay into two $\phi_{0}$ bosons.
This new decay channel can change other decay branching ratios and
thus affect the strategy of
searching for the Higgs boson at high energy colliders, which is the main aim of this paper.\\
\indent Ref.\cite{1003} studied the Higgs phenomenology in LRTH
model by paying special attention to the decay $h\rightarrow
\hat{S}\hat{S}$ which is strongly corrected with the dark matter
scattering on nucleon. They found that such an invisible decay can
severely suppress the conventional decay modes like $h\rightarrow
VV(V=W, Z)$ and $h\rightarrow b\bar{b}$. Note that similar exotic
decays for the SM-like Higgs boson may also be predicted by some
other new physics models like the little Higgs models and SUSY or
two Higgs-doublet models et al \cite{logan,susy}. A common feature
 of their phenomenology is the suppression of the conventional visible channels of the Higgs boson.
 To distinguish between different models, all the channels of Higgs production should be jointly analyzed.
 In this work we first study the decay branching ratios
of the Higgs boson in the LRTH model for small value of
$m_{\phi_{0}}$. Then we study the production via gluon fusion
followed by the decay into two photons or two charged gauge bosons
in the cases of $m_{h}>2m_{\phi_{0}}$ and $m_{h}<2m_{\phi_{0}}$,
respectively. We also study the process
$\gamma\gamma\rightarrow h \rightarrow b\bar{b}$ at ILC for these two cases.\\
\indent This article is organized is as follows. In the next
section, we briefly review the left-right twin Higgs model. In Sec.
III, we calculate the decay branching ratios of the Higgs boson. In
Sec. IV, we calculate the main production of the Higgs boson at the
LHC via gluon fusion followed by the decay into two photons or two
weak gauge bosons. In Sec. V we calculated the rate of
$\gamma\gamma\rightarrow h \rightarrow b\bar{b}$ at ILC. Finally, we
give our conclusion in Sec.VI.\\
 \noindent{\bf II. Review of the left-right twin Higgs model}\\
\indent Before our calculations we recapitulate the left-right twin
Higgs (LRTH) model. The details of the LRTH model as well as the
particle spectrum, Feynman rules, and some phenomenology analysis
have been studied in Ref.\cite{Hock}. Here we will briefly review
the essential features of the LRTH model and focusing on the new
particles and the
couplings relevant to our computation.\\
\indent The LRTH model is based on the global $U(4)_{1}\times
U(4)_{2}$ symmetry with a locally gauged subgroup $SU(2)_{L}\times
SU(2)_{R}\times U(1)_{B-L}$. The twin symmetry is identified with
the left-right symmetry which interchanges L and R, implying that
the gauge couplings of $SU(2)_{L}$ and $SU(2)_{R}$ are identical
$(g_{2L}=g_{2R}=g_{2})$. Two Higgs fields, $H$ and $\hat{H}$, are
introduced and each transforms as $(4,1)$ and $(1,4)$ respectively
under the global symmetry. They can be written as
\begin{eqnarray}
H=\left( \begin{array}{c} H_{L}\\
H_{R} \\
\end{array}  \right)\,,~~~~~~~~~~~~~~\hat{H}=\left( \begin{array}{c} \hat{H}_{L}\\
\hat{H}_{R} \\
\end{array}  \right)\,,
\end{eqnarray}
where $H_{L,R}$ and $\hat{H}_{L,R}$ are two component objects which
are charged under the $SU(2)_{L}\times SU(2)_{R}\times U(1)_{B-L}$
as
\begin{equation}
H_{L}~and~ \hat{H}_{L}: (2, 1, 1),~~~~~~~~H_{R}~ and~ \hat{H}_{R}:
(1, 2, 1).
\end{equation}
The global $U(4)_{1}(U(4)_{2})$ symmetry is spontaneously broken
down to its subgroup $U(3)_{1}(U(3)_{2})$ with non-zero vacuum
expectation values(VEV) as $\langle H\rangle=(0,0,0,f)$ and $\langle
\hat{H}\rangle=(0,0,0,\hat{f})$. Each spontaneously symmetry
breaking results in seven Nambu-Goldstone bosons. Three of six
Goldstone bosons that are charged under $SU(2)_{R}$ are eaten by the
new gauge bosons $W_{H}^{\pm}$ and $Z_{H}$, while leaves three
physical Higgs: $\phi^{0}$ and $\phi^{\pm}$. After the SM
electroweak symmetry breaking, the three additional Goldstone bosons
are eaten by the SM gauge bosons $W^{\pm}$ and $Z$. The remaining
Higgses are the SM Higgs doublet $H_{L}$ and an extra Higgs doublet
$\hat{H}_{L}=(\hat{H}_{1}^{+},\hat{H}_{2}^{0})$ that only couples to
the gauge boson sector. A residue matter parity in the model renders
the neutral Higgs $\hat{H}_{2}^{0}$ stable, and it could be a good
dark matter candidate. \\
\indent In the LRTH model, the masse of charged gauge bosons and
fermions are given by \cite{Hock}
\begin{eqnarray}
m_{W_{L}}^{2}&=& \frac{1}{2}g_{2}^{2}f^{2}\sin^{2}x, \\
m_{W_{H}}^{2}&=& \frac{1}{2}g_{2}^{2}(\hat{f}^{2}+f^{2}\cos^{2}x), \\
m_{t}^{2}&=& \frac{1}{2}(M^{2}+y^{2}f^{2}-N_{t}), \\
m_{T}^{2}&=& \frac{1}{2}(M^{2}+y^{2}f^{2}+N_{t}),
\end{eqnarray}
where $N_{t}=\sqrt{(M^{2}+y^{2}f^{2})^{2}-y^{4}f^{4}\sin^{2}2x}$
with $x=v/\sqrt{2}f$, in which $v=246GeV$ is the electroweak scale.
$g_{2}=e/\sin\theta_{W}$ and $\theta_{W}$ is the Weinberg angle. The
values of $f$ and $\hat{f}$ will be bounded by electroweak precision
measurements. Once $f$ is fixed, the values of $\hat{f}$ can be
determined from the minimization of the Coleman-Weinberg potential
of the SM Higgs. The mass parameter $M$ is essential to the mixing
between the SM-like top quark and the heavy T-quark. \\
\indent At the leading order, the couplings expression forms of the
Higgs boson with charged gauge bosons and fermions, which are
related to our calculation can be written as \cite{Hock}
\begin{eqnarray}
hWW &:& \frac{1}{2}g_{2}^{2}v(1-\frac{v^{2}}{3f^{2}}), ~~~~~~~~~~
hW_{H}W_{H}: -\frac{1}{2}g_{2}^{2}v(1-\frac{v^{2}}{3f^{2}}), \\
ht\bar{t} &:& -\frac{m_{t}}{v}C_{L}C_{R}, ~~~~~~~~~~~~~~hT\bar{T}:
-\frac{y}{\sqrt{2}}(S_{R}S_{L}-C_{L}C_{R}x),
\end{eqnarray}
where
\begin{eqnarray}
S_{L}&=& \frac{1}{\sqrt{2}}\sqrt{1-(y^{2}f^{2}\cos2x+M^{2})/N_{t}},~~~~~C_{L}=\sqrt{1-S_{L}},\\
S_{R}&=&
\frac{1}{\sqrt{2}}\sqrt{1-(y^{2}f^{2}\cos2x-M^{2})/N_{t}},~~~~~C_{R}=\sqrt{1-S_{R}}.
\end{eqnarray}
\indent The Coleman-Weinberg potential, obtained by intefrating out
the gauge bosons and top quarks, yields the SM Higgs potential,
which determine the SM Higgs VEV and its mass, as well as the masses
for the other Higgs. On the other hand,  the $\mu$-term,
 \begin{eqnarray}
V_{\mu}=-\mu_{r}^{2}(H_{R}^{\dagger}\hat{H}_{R}+h.c.)+\hat{\mu}^{2}H_{L}^{\dagger}\hat{H}_{L},
 \end{eqnarray}
 contributes to the Higgs masses at tree level. The mass of
 $\phi_{0}$ and new scalar self-interactions are given by \cite{Hock}
 \begin{eqnarray}
m^{2}_{\phi_{0}}&=&\frac{\mu_{r}^{2}f\hat{f}}{\hat{f}^{2}+f^{2}\cos^{2}x}[\frac{\hat{f}^{2}(\cos
x+\frac{\sin x}{x}(3+x^{2}))}{f^{2}(\cos x+\frac{\sin
x}{x})^{2}}+2\cos
x+\frac{f^{2}\cos^{2}x(1+\cos x)}{2\hat{f}^{2}}],\\
h\phi_{0}\phi_{0} &:& x(30p2\cdot p3+11p1\cdot p1)/(27\sqrt{s}f)),
 \end{eqnarray}
here $p1$, $p2$, and $p3$ refer to the incoming momentum of the
first, second, and third particle, respectively. From above we can
see that the mass of the neutral Higgs boson $\phi_{0}$ is a free
parameter and is determined by $\mu_{r}$ and $f$. Here we consider
another possibility, in which the mass is in the low mass region
where the new decay $h\rightarrow \phi_{0}\phi_{0}$ can be
open.\\
\noindent{\bf III. Higgs decay branching ratios in LRTH model}~~\\
 \indent  In the LRTH model, the major decay modes of the Higgs boson are the SM-like ones:
 $h\rightarrow f\bar{f}$($f=b,c,\tau$), $WW$ and $ZZ$. The LRTH model gives corrections to these decay modes via the corresponding
 modified Higgs couplings
 \begin{eqnarray}
\Gamma(h\rightarrow XX)=\Gamma(h\rightarrow
XX)_{SM}(g_{hXX}/g_{hXX}^{SM})^{2},
 \end{eqnarray}
 where $X$ denotes a SM particle, $\Gamma(h\rightarrow
XX)_{SM}$ is the decay width in the SM, and the $g_{hXX}$ and
$g_{hXX}^{SM}$ are the couplings of $hXX$ in the LRTH model and SM,
respectively. The loop-induced decays $H\rightarrow gg$ and
$H\rightarrow \gamma\gamma$ will be also important for a low Higgs
mass. In the LRTH model, in addition to the corrections via the
modified couplings $ht\bar{t}$ and $hWW$, the new heavy T-quark and
charged gauge bosons can also contribute to their decay widths
\cite{epl}. For the decay $h\rightarrow Z\gamma$,
 the $W$ boson loop contribution is dominant\cite{phys} and thus we only
 consider the alteration of the Higgs coupling with the $W$ boson.
 Because the QCD radiative corrections are rather small\cite{qcd}, our results
 is precise enough.\\
 \indent As discussed in \cite{loop}, the mass of neutral Higgs boson
 $\phi_{0}$ may be as low as $50 GeV$. Therefore, in addition to the SM decay channels,
 the new decay $h\rightarrow
\phi_{0}\phi_{0}$ will open for $m_{h}\geq 2m_{\phi_{0}}$, and the
partial width is given by
 \begin{eqnarray}
\Gamma(h\rightarrow
\phi_{0}\phi_{0})&=&\frac{g^{2}_{h\phi_{0}\phi_{0}}}{8\pi
m_{h}}\sqrt{1-\frac{4m_{\phi}^{2}}{m_{h}^{2}}},
 \end{eqnarray}
here $g_{h\phi_{0}\phi_{0}}$ is the couplings
of $h\phi_{0}\phi_{0}$.\\
\indent In the LRTH model, the SM-like Higgs mass can be obtained
via the minimization of the Higgs potential, which depends slightly
on $M$ and $\Lambda$ but is insensitive to $f$. Varying $M$ between
0 and $150 GeV$, $\Lambda$ between $2\pi f$ and $4\pi f$, its mass
is found to be in the range of $145-180 GeV$ \cite{Hock}. However,
if we take a little smaller value of $\Lambda$, the lower bound on
the SM-like Higgs mass will be relaxed. In our calculations, the
free parameters involved are $f$, $M$, $m_{\phi_{0}}$ and the Higgs
boson mass $m_{h}$. For the Higgs boson mass, we will take it in the
range of $110 GeV-200 GeV$. Following Ref.\cite{Hock}, we will
assume that the values of the free parameters $f$ and $M$ are in the
ranges of $500 GeV- 1500 GeV$ and $0\leq M\leq f$, respectively.
 \begin{figure}[ht]
\begin{center}
\scalebox{0.8}{\epsfig{file=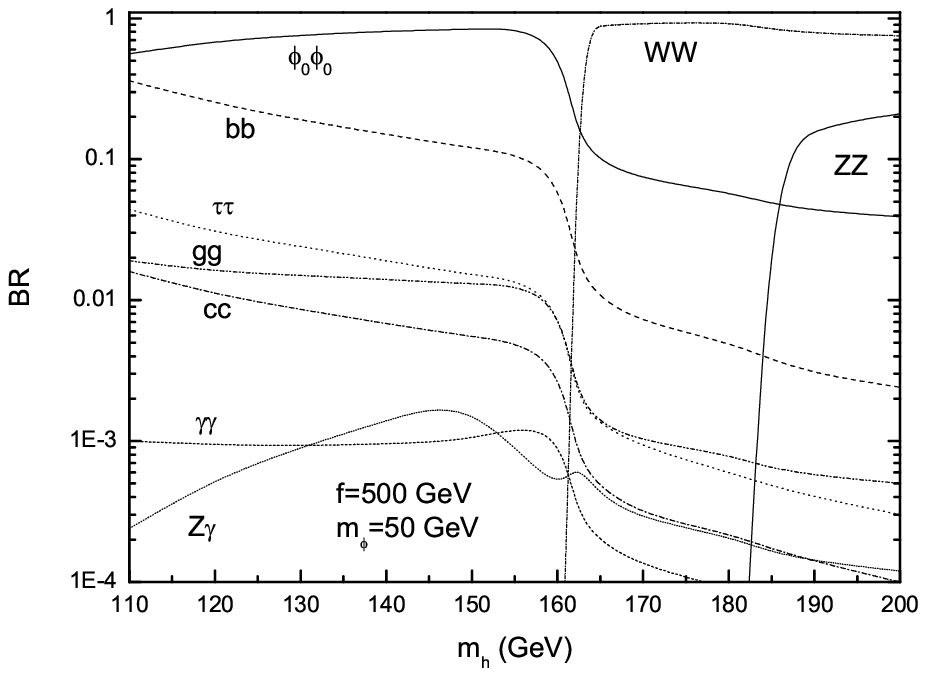}\epsfig{file=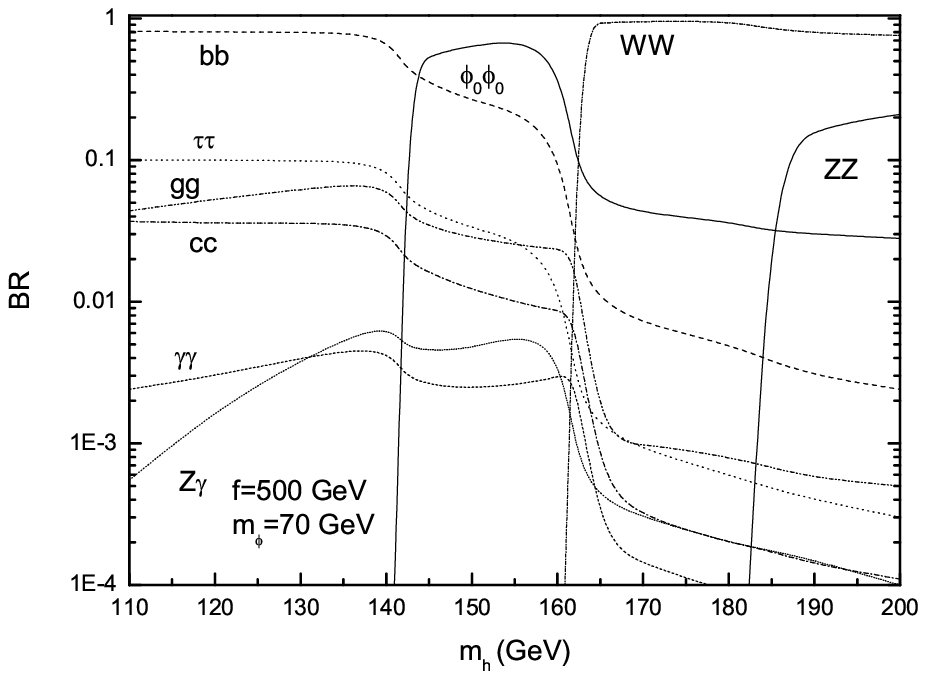}}\\
\end{center}
\caption{\small The Higgs decay branching ratios versus the Higgs
mass in LRTH model.}
\end{figure}\\
\indent A search strategy of the Higgs boson depends sensitively on
its branching ratios(BR): In the SM, the major decay mode for
$m_{h}< 2m_{W}$ is into $b\bar{b}$ while that for $m_{h}> 2m_{W}$ is
into $W^{+}W^{-}$. In the LRTH model, there may be one new decay
mode $h\rightarrow \phi_{0}\phi_{0}$ for Higgs boson. Fig. 1 show
the Higgs deay branching ratios as a function of the Higgs mass
$m_{h}$ in the LRTH model for $f=500 GeV$ and $m_{\phi_{0}}=50 GeV$,
$70 GeV$, respectively. We see that the dominant decay channel is
$h\rightarrow WW$ for $160 GeV< m_{h}<200 GeV$, similar to the SM
prediction. But for the $m_{\phi_{0}}=50 GeV$ case, the decay
$h\rightarrow \phi_{0}\phi_{0}$ is dominant and over $70\%$ for $120
GeV< m_{h}< 160 GeV$, then it decreases as $m_{h}$ gets large and
become comparable with $h\rightarrow WW$ at about 160 GeV. For
$2m_{\phi_{0}}<m_{h}<160 GeV$, the decay width of $h\rightarrow
\phi_{0}\phi_{0}$ is much larger than the decay $h\rightarrow
b\bar{b}$. The reason is that the Higgs couplings is of the
electroweak strength and much larger than the Yukawa coupling of $b$
quark. Here we fixed $f=500 GeV$ and did not show the dependence of
$f$. the decay $h\rightarrow \phi_{0}\phi_{0}$  becomes less
important as $f$ gets larger.\\
\indent In table 1, we list the Higgs decay branching ratios
normalized to the SM predictions for three main channels in the LRTH
model. Table 1 shows that the deviation from the SM prediction for
each decay mode becomes small as $f$ gets large. The deviation from
the SM prediction is also sensitive to the Higgs boson mass. For
$m_{h}=120 GeV$, and $500 GeV\leq f\leq 1000 GeV$, the deviations
for the
decay $h\rightarrow b\bar{b}$ and $h\rightarrow gg$ are in the ranges of $11\%-68\%$, $18\%-78\%$, respectively.
For the decay modes $h\rightarrow gg$ and $h\rightarrow \gamma\gamma$, the deviations from the SM
predictions are also sensitive to $M$, which are not shown here.
 We will show the dependence of these decay modes on $M$ later.\\
\null\noindent ~~{\bf Table 1:} The value
$R_{BR}=BR^{LRTH}(h\rightarrow XX)/BR^{SM}(h\rightarrow XX)$ is
defined as a ratio of the Higgs decay branching ratios in LRTH model
to one in the SM for $m_{h}=(120, 150, 180)GeV$, where $X=b\bar{b}$,
$gg$ and $\gamma\gamma$ with $M=150 GeV$ and $m_{\phi_{0}}=50 GeV$ .
\vspace{0.1in}
\begin{center}
\doublerulesep 0.8pt \tabcolsep 0.1in
\begin{tabular}{||c|c|c|c|c|c|c|c|c|c||}\hline\hline
f (GeV)&\multicolumn{3}{|c|}{$R_{BR}(b\bar{b})$}&\multicolumn{3}{|c|}{$R_{BR}(gg)$}&\multicolumn{3}{|c|}{$R_{BR}(\gamma\gamma)$}\\
\hline 500&0.32&0.17&0.94&0.22&0.11&0.66&0.31&0.16&0.89\\
\hline 700&0.65&0.44&0.98&0.54&0.37&0.83&0.64&0.43&0.96\\
\hline 1000&0.89&0.77&0.99&0.82&0.71&0.92&0.88&0.76&0.99\\
\hline
\end{tabular}
\end {center}
\vspace{0.5cm}

 \noindent{\bf IV. The rates $\sigma(gg
\rightarrow h)\times
BR(h\rightarrow \gamma\gamma(W^{+}W^{-}))$ at LHC in the LRTH model}~~\\
\indent In the SM the Higgs production at the LHC is dominated by
gluon fusion process. The $h\rightarrow\gamma\gamma$ channel shows
very good sensitivity in the range of $114 GeV< m_{h}<140 GeV$.
Especially, the rate $\sigma(gg \rightarrow h)\times BR(h\rightarrow
\gamma\gamma)$ can be measured to $10\%(30\%)$ with an integrated
luminosity $100 fb^{-1}(10 fb^{-1})$ from both ATLAS and
CMS\cite{atlas}. Once we find a light Higgs boson at the LHC, this
channel can provide a test for different models. In the LRTH model,
$\sigma(gg\rightarrow h)$ is strongly correlated with  the decay
width $\Gamma(gg\rightarrow h)$. In our results we use
$\sigma(gg\rightarrow h)$ to denote the hadronic cross section of
the Higgs production proceeding through $gg\rightarrow h$ at parton
level. We use CTEQ6L \cite{cteq} for parton distributions, with the
renormalization scale $\mu_{R}$ factorization scale $\mu_{F}$ chosen
to be
$\mu_{R}=\mu_{R}=m_{h}$.\\
%%%%%%%%%%%%%%%%%%%%%%%%%%%%%%%%%%%%%%%%%%%%%%%%%%%%%%%%%%%%%%%%%
\begin{figure}[t]
\begin{center}
\scalebox{0.85}{\epsfig{file=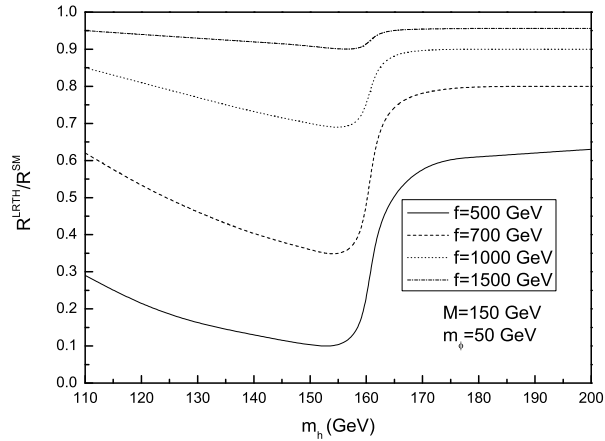}}\\
\end{center}
\caption{\small The value of $\sigma(gg \rightarrow h)\times
BR(h\rightarrow \gamma\gamma)$ normalized to the SM prediction in
the LRTH model as a function of Higgs boson mass for $M=150 GeV$ and
various values of $f$ as indicated.}
\end{figure}
\begin{figure}[t]
\begin{center}
\scalebox{0.85}{\epsfig{file=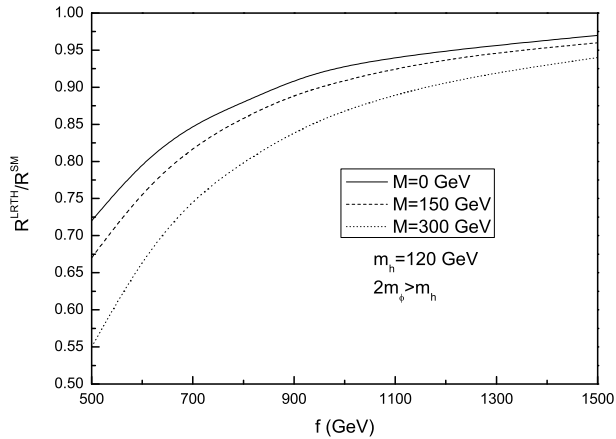}}\\
\end{center}
\caption{\small Same as Fig.2 but for the case of
$m_{h}<2m_{\phi_{0}}$. }
\end{figure}
\indent In Fig. 2 show the rates of $\sigma(gg \rightarrow h)\times
BR(h\rightarrow \gamma\gamma)$ normalized to the SM prediction in
the LRTH model as a function of $m_{h}$ for $M=150 GeV$,
$m_{\phi_{0}}=50 GeV$ and various values of $f$. One can see from
Fig. 2 that compared with the SM predictions, the LRTH model can
suppress the rates sizably for a small value of $f$. The reason for
such a severe suppression is that the decay mode $h\rightarrow
\phi_{0}\phi_{0}$ can be dominant in some part of the parameter
space and thus the total decay width of Higgs boson becomes much
larger than the SM value. For example, for $f=500 GeV$ and
$m_{H}=120(150) GeV$, the rates are suppressed to
about 0.2(0.1) relative to the SM predictions in LRTH model. \\
\indent Fig. 3 show the rates of $\sigma(gg \rightarrow h)\times
BR(h\rightarrow \gamma\gamma)$ normalized to the SM prediction in
the LRTH model as a function of $m_{h}$ for the case of
$m_{h}<2m_{\phi_{0}}$. One can see that, as $f$ gets large, the
suppression is weakened sharply. The deviation from the SM
prediction is also sensitive to the mixing parameter $M$. This is
because $M$ is introduced to generate the mass mixing term
$Mq_{L}q_{R}$, and the LRTH model can give corrections via the
coupling of $ht\bar{t}$ and the heavy T-quark loop. For $m_{h}=120
GeV$ and $f=500 GeV$, the suppression of SM predictions can reach
$28\%$ and $33\%$ for $M=0 GeV$ and $M=150 GeV$, respectively. The
rate for $gg\rightarrow H\rightarrow \gamma\gamma$ can be measured
with a precision of $10-15\%$ for $m_{h}<150 GeV$.
 Therefore, it is possible to probe the LRTH model via such a process at the LHC. \\
\begin{figure}[b]
\begin{center}
\scalebox{0.85}{\epsfig{file=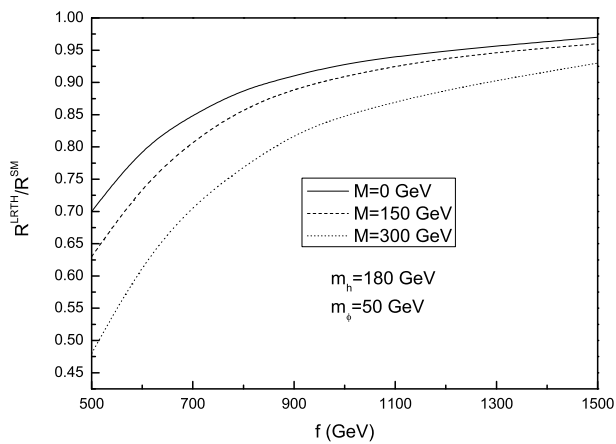}}\\
\end{center}
\caption{\small The value of $\sigma(gg \rightarrow h)\times
BR(h\rightarrow W^{+}W^{-})$ normalized to the SM prediction in the
LRTH model versus the value of $f$ with three values of $M$ as
indicated. }
\end{figure}
\indent When Higgs mass is relatively heavy ($2m_{W}< m_{h}<
2m_{Z}$), the decay $h\rightarrow WW\rightarrow l\nu l\nu$ is an
excellent channel for searching for Higgs boson \cite{hww}. In Fig.
4 we plot the rates of $\sigma(gg \rightarrow h)\times
BR(h\rightarrow W^{+}W^{-})$ normalized to the SM prediction in the
LRTH model versus the value of $f$ for $m_{h}=180 GeV$ and
$m_{\phi_{0}}=50 GeV$. We see that, compared with the SM prediction,
the LRTH model can suppress the rates significantly for a small of
$f$. For $M=150 GeV$ and $500 GeV\leq f \leq 800 GeV$,
the suppression of SM prediction is in the range of $37\%-14\%$,
 which can exceed the experimental uncertainty ($10\%-20\%$)\cite{exp}.\\
\indent It has been shown \cite{0611242} that, the Higgs boson can
dominantly decay into a pair of pseudoscalar boson. Together with
smaller $g_{ZZh}$ and $B(h\rightarrow b\bar{b})$ than in the SM, the
LEP Higgs boson mass bound based on the limit
$(g_{ZZh}/g_{ZZh}^{SM})^{2}B(h\rightarrow b\bar{b})$ can be reduced.
In Ref.\cite{prl}, the authors shown that $h\rightarrow
\eta\eta\rightarrow b\bar{b}b\bar{b}$ is complementary and can be
used to detect the intermediate Higgs boson at the LHC, via $Wh$ and
$Zh$ production. In the LRTH model, $\phi_{0}$ mainly decays into
$b\bar{b}$, $c\bar{c}$, or $\tau^{+}\tau^{-}$. The decay branching
ratio of $\phi_{0}\rightarrow b\bar{b}$, $c\bar{c}$, and
$\tau^{+}\tau^{-}$ are close to the corresponding SM Higgs decay
branching ratios \cite{Hock}. Thus the ultimate dominant decay mode
of the Higgs can be $h\rightarrow \phi_{0}\phi_{0}\rightarrow
b\bar{b}b\bar{b}$. Detailed study needs to be done to optimize the
cuts and identify the signal from the background. Such study is
beyond the scope of the current paper and we leave it for future
work.\\
 \noindent{\bf V. The process $\gamma\gamma \rightarrow
h\rightarrow
b\bar{b}$ in the LRTH model}~~\\
\indent While the LHC is widely regarded as discovery machine for
Higgs boson, a precision measurement of Higgs property can be only
achieved at the proposed International Linear Collider
(ILC)\cite{ILC}. A unique feature of the ILC is that it can be
transformed to $\gamma\gamma$ modes by the laser-scattering method.
Such an option of photon-photon collision can possibly measure the
rates of the Higgs production with a precision of a few percent.
Especially, for $\gamma\gamma\rightarrow h \rightarrow b\bar{b}$
process, the production rate could be measured at about $2\%$ for a
light Higgs boson \cite{bb}. Such a process $\gamma\gamma
\rightarrow h\rightarrow b\bar{b}$ is a sensitive probe for new
physics because the loop-induced $h\gamma\gamma$ coupling and the
$hb\bar{b}$ coupling
are sensitive to new physics \cite{jhep}. \\
%%%%%%%%%%%%%%%%%%%%%%%%%%%%%%%%%%%%%%%%%%%%%%%%%%%%%%%%%%%%%%%%%
\begin{figure}[ht]
\begin{center}
\scalebox{0.85}{\epsfig{file=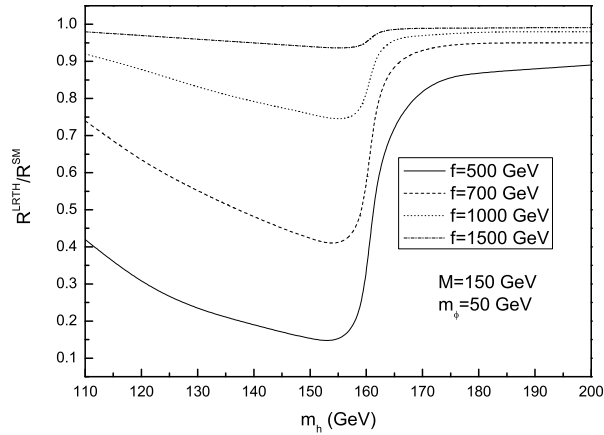}}\\
\end{center}
\caption{\small The value of $\sigma(\gamma\gamma\rightarrow
h)\times BR(h\rightarrow b\bar{b})$ normalized to the SM prediction
in the LRTH model as a function of $m_{h}$ for $M=150 GeV$,
$m_{\phi_{0}}=50 GeV$, and various values of $f$ as indicated.}
\end{figure}
%%%%%%%%%%%%%%%%%%%%%%%%%%%%%%%%%%%%%%%%%%%%%%%%%%%%%%%%%%%%%%%%%
\begin{figure}[ht]
\begin{center}
\scalebox{0.85}{\epsfig{file=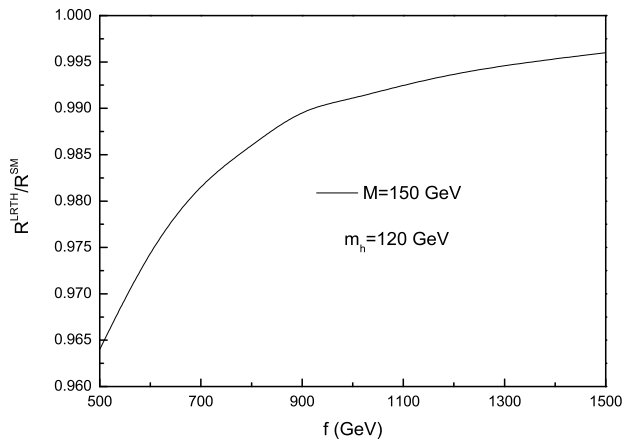}}\\
\end{center}
\caption{\small Same as Fig. 5 but for the case of
$m_{h}<2m_{\phi_{0}}$.}
\end{figure}
\indent Figures 5 and 6 show the numerical results for the rate
$\sigma(\gamma\gamma\rightarrow h)\times BR(h\rightarrow b\bar{b})$
normalized to the SM prediction in the LRTH model, for
$m_{\phi_{0}}=50 GeV$, and $m_{h}<2m_{\phi_{0}}$, respectively. From
Fig. 5, we can see that the rate have a sizable deviation from the
SM prediction, and the magnitude of deviation is sensitive to the
scale $f$. For $M=150 GeV$, $m_{h}=120 GeV$, and $500 GeV\leq f\leq
1000 GeV$, the suppression is in the range of $70\%-12\%$. The
reason for such a serve suppression is similar to what have been
discussed above, i.e., the opening of new decay mode. In the case of
$m_{h}<2m_{\phi_{0}}$, the new decay mode $h\rightarrow
\phi_{0}\phi_{0}$ is kinematically forbidden. Fig.6 show that the
LRTH model also suppresses the rate $\sigma(\gamma\gamma\rightarrow
h)\times BR(h\rightarrow b\bar{b})$, but the suppression can only
reach about $4\%$. This is because the contribution from the LRTH
model mainly come from the loops of new top partner and heavy
charged gauge boson, in addition to the modified couplings
$ht\bar{t}$ and $hWW$ at order $v^{2}/f^{2}$\cite{epl}. For a large value of $f$, the suppression is only a few percent. \\
\noindent{\bf VI. Conclusion}~~\\
\indent The twin Higgs mechanism provides an alternative method to
solve the little hierarchy problem. The Left-right twin Higgs model
is a concrete realization of the twin Higgs mechanism, which
predicts one neutral scalar particle $\phi_{0}$. With the $\mu$ term
introduced by hand, the $\phi_{0}$ boson acquires mass
$m_{\phi_{0}}\sim \mu_{r}$, which can be lighter than half the Higgs
boson mass in a portion of parameter space. In this paper we focus
on the case of $m_{h}\geq 2m_{\phi_{0}}$ so that the new decay
$h\rightarrow \phi_{0}\phi_{0}$ can be open. From our numerical
results we obtain the following observations: (i) For the Higgs
decay, we found that, with $f=500 GeV$ and $2m_{\phi_{0}}<m_{h}<160
GeV$, the new decay $h\rightarrow \phi_{0}\phi_{0}$ can be the
dominant mode and it can give very different branching ratios from
the SM prediction. The branching ratios of the conventional decay
modes of the Higgs boson, $h\rightarrow gg$ and $h\rightarrow
b\bar{b}$, can be suppressed over $60\%$, $50\%$, respectively;
(ii)For the rates $\sigma(gg \rightarrow h)\times BR(h\rightarrow
\gamma\gamma(W^{+}W^{-}))$ at the LHC, the LRTH model can give
severe suppression relative to the SM predictions, whenever the
neutral scalar mass is less than the mass of the Higgs boson; (iii)
For the process $\gamma\gamma\rightarrow h \rightarrow b\bar{b}$,
the LRTH model can always suppress the rate for the cases of
$m_{h}>2m_{\phi_{0}}$ and $m_{h}<2m_{\phi_{0}}$, respectively.
However, the production rate can be severely suppressed in some of
the parameter space where the new decay mode is open and dominant
for the case of $m_{h}>2m_{\phi_{0}}$; (iv) The Higgs production
cross section times the branching ratios of the conventional decays
can be all suppressed significantly for a small value of the scale
$f$. Therefore, it is possible to probe the LRTH model via these
Higgs boson production processes at the LHC
experiment or in the future ILC experiment.\\
\indent  \\
 \vspace{.5cm}
\noindent{\bf Acknowledgments}

 \indent This work is supported in
part by the National Natural Science Foundation of China(Grant
No.10775039).
 
\end{document}